\documentclass[12pt]{article}%
\usepackage{amsfonts}
\usepackage{amsmath}
\usepackage{amssymb}
\usepackage{graphicx}%
\providecommand{\U}[1]{\protect\rule{.1in}{.1in}}
\newtheorem{theorem}{Theorem}

\newtheorem{notation}[theorem]{Notation}

\begin{document}

\title{On the Fermi Function of Squeezed Coherent States}
\author{Maurice A. de Gosson\thanks{This work has been supported by the Austrian
Research Agency FWF (Projektnummer P20442-N13).}\\\ \textit{University of Vienna }\\\textit{Faculty of Mathematics, NuHAG}\\\textit{A-1090 Vienna, AUSTRIA}}
\maketitle

\begin{abstract}
Fermi observed in 1930 that the state of a quantum system may be defined in
two different (but equivalent) ways, namely by its wavefunction $\Psi$ or by a
certain function $g_{\mathrm{F}}$ on phase space canonically associated with
$\Psi$. In this Note we study we study Fermi's function when $\Psi$ is a
squeezed coherent state. We relate it with the Wigner transform of $\Psi$,
thus generalising a previous observation of Benenti and Strini. We show that
the symplectic capacity of the phase space ellipsoid $g_{\mathrm{F}}%
(x,p)\leq0$ is bounded by $h/2$ and $nh/2$ ($n$ the number of degrees of
freedom).  

\end{abstract}

\section{Introduction \label{secfe}}

In a largely forgotten paper \cite{Fermi} from 1930 Enrico Fermi associates to
every quantum state $\Psi$ a certain hypersurface $g_{\mathrm{F}}(x,p)=0$ in
phase space. The underlying idea is actually extremely simple, and consists in
observing that any complex twice continuously differentiable function
$\Psi(x)=R(x)e^{i\Phi(x)/\hslash}$ ($R(x)\geq0$ and $\Phi(x)$ real) defined on
$\mathbb{R}^{n}$ satisfies the partial differential equation%
\begin{equation}
\left[  \left(  -i\hbar\nabla_{x}-\nabla_{x}\Phi\right)  ^{2}+\hbar^{2}%
\frac{\nabla_{x}^{2}R}{R}\right]  \Psi=0.\label{gf1}%
\end{equation}
where $\nabla_{x}^{2}$ is the Laplace operator in the variables $x_{1}%
,...,x_{n}$. Performing the gauge transformation $-i\hbar\nabla_{x}%
\longrightarrow-i\hbar\nabla_{x}-\nabla_{x}\Phi$ this equation is in fact
equivalent to the trivial equation%
\begin{equation}
\left(  -\hbar^{2}\nabla_{x}^{2}+\hbar^{2}\frac{\nabla_{x}^{2}R}{R}\right)
R=0.\label{trivial}%
\end{equation}
The operator
\begin{equation}
\widehat{g_{\mathrm{F}}}=\left(  -i\hbar\nabla_{x}-\nabla_{x}\Phi\right)
^{2}+\hbar^{2}\frac{\nabla_{x}^{2}R}{R}\label{fermop}%
\end{equation}
appearing in the left-hand side of Eqn. (\ref{gf1}) is the quantisation (in
every reasonable physical quantisation scheme) of the real observable
\begin{equation}
g_{\mathrm{F}}(x,p)=\left(  p-\nabla_{x}\Phi\right)  ^{2}+\hbar^{2}%
\frac{\nabla_{x}^{2}R}{R}\label{gf2}%
\end{equation}
and the equation $g_{\mathrm{F}}(x,p)=0$ determines a hypersurface
$\mathcal{H}_{\mathrm{F}}$ in phase space $\mathbb{R}_{x,p}^{2n}$ which Fermi
ultimately \emph{identifies} with the state $\Psi$ itself.

Of course, Fermi's analysis was very heuristic and its mathematical rigour
borders the unacceptable (at least by modern standards). Fermi's paper has
recently been rediscovered by Benenti \cite{benenti} and Benenti and Strini
\cite{best}, who study its relationship with the level sets of the Wigner
transform of $\Psi$. 

The aim of the present Note is to push further the analysis in
\cite{benenti,best} by considering the Fermi function of squeezed coherent
states. We will also study the symplectic capacity of the corresponding set
bounded by the Fermi surface $\mathcal{H}_{\mathrm{F}}$ (which is in this case
an ellipsoid).

\begin{notation}
The points in configuration and momentum space are written $x=(x_{1}%
,...,x_{n})$ and $p=(p_{1},...,p_{n})$; in formulas $x$ an $p$ are viewed as
column vectors. We will also use the collective notation $z=(x,p)$ for the
phase space variable. The matrix $J=%
\begin{bmatrix}
0 & I\\
-I & 0
\end{bmatrix}
$ ($0$ and $I$ the $n\times n$ zero and identity matrices) defines the
standard symplectic form on the phase space $\mathbb{R}_{x}^{2n}$ via the
formula $\sigma(z,z^{\prime})=Jz\cdot z^{\prime}=p\cdot x^{\prime}-p^{\prime
}\cdot x$. We write $\hbar=h/2\pi$, $h$ being Planck's constant.
\end{notation}

\section{The Harmonic Oscillator}

As an appetizer we begin by considering the fiducial coherent state
\begin{equation}
\Psi_{0}(x)=\left(  \tfrac{1}{\pi\hbar}\right)  ^{n/4}e^{-|x|^{2}/2\hbar}
\label{fid}%
\end{equation}
with $|x|^{2}=x\cdot x$; it is the ground state of the $n$-dimensional
isotropic harmonic oscillator with mass and frequency equal to one:%
\begin{equation}
\tfrac{1}{2}(-\hbar^{2}\nabla_{x}^{2}+|x|^{2})\Psi_{0}=\tfrac{1}{2}n\hbar
\Psi_{0}. \label{gf5}%
\end{equation}
The operator (\ref{gf1}) is here%
\begin{equation}
\widehat{g_{\mathrm{F}}}=-\hbar^{2}\nabla_{x}^{2}+|x|^{2}-n\hbar\label{gf4}%
\end{equation}
and the relation $\widehat{g_{\mathrm{F}}}\Psi_{0}=0$ is hence equivalent to
Eqn. (\ref{gf5}). The Fermi function is
\begin{equation}
g_{\mathrm{F}}(x,p)=|p|^{2}+|x|^{2}-n\hbar\label{gf7}%
\end{equation}
and the ellipsoid $\mathcal{W}_{\mathrm{F}}$ is thus the disk $|x|^{2}%
+|p|^{2}\leq n\hbar$ whose area is $n\pi\hbar=nh/2$.

Consider next the $N$-th eigenstate $\Psi_{N}$; assume first $n=1$. We have
\[
\frac{1}{2}\left(  -\hbar^{2}\frac{d^{2}}{dx^{2}}+x^{2}\right)  \Psi
_{N}=\left(  N+\frac{1}{2}\right)  \hbar
\]
and the eigenfunction $\Psi_{N}$ is the (unnormalized) Hermite function%
\begin{equation}
\Psi_{N}(x)=e^{-|x|^{2}/2\hbar}H_{N}(x/\sqrt{\hbar}) \label{hermite1}%
\end{equation}
where
\[
H_{N}(x)=(-1)^{n}e^{x^{2}}\frac{d^{N}}{dx^{N}}e^{-x^{2}}%
\]
is the $N$-th Hermite polynomial. Since $\Psi_{N}$ is real, the corresponding
Fermi function is
\begin{equation}
g_{\mathrm{F}}(x,p)=p^{2}+x^{2}-(2N+1)\hbar\label{gf6}%
\end{equation}
and the Fermi set $g_{\mathrm{F}}(x,p)$ is the circle%
\begin{equation}
p^{2}+x^{2}=(2N+1)\hbar\label{ball1}%
\end{equation}
whose area is $(2N+1)\pi\hbar=\left(  N+\frac{1}{2}\right)  h$. In the case of
an arbitrary number $n$ of degrees of freedom the eigenstate $\Psi_{N}$ is the
tensor product of $n$ Hermite functions (\ref{hermite1}) and one finds that
\[
g_{\mathrm{F}}(x,p)=|p|^{2}+|x|^{2}-(2N+1)\hbar
\]
hence the Fermi set is this time the ball%
\begin{equation}
|p|^{2}+|x|^{2}=(2N+1)\hbar. \label{ball2}%
\end{equation}

\section{Squeezed Coherent States}

We next consider arbitrary (normalized) squeezed coherent states%
\begin{equation}
\Psi_{X,Y}(x)=\left(  \frac{1}{\pi\hbar}\right)  ^{n/4}(\det X)^{1/4}%
\exp\left[  -\frac{1}{2\hbar}(X+iY)x\cdot x\right]  \label{coh1}%
\end{equation}
where $X$ and $Y$ are real symmetric $n\times n$ matrices, and $X$ is positive
definite. Setting $\Phi(x)=-\frac{1}{2}Yx\cdot x$ and $R(x)=\exp\left(
-\frac{1}{2\hbar}Xx\cdot x\right)  $ we have
\begin{equation}
\nabla_{x}\Phi(x)=-Yx\text{ \ , \ }\frac{\nabla_{x}^{2}R(x)}{R(x)}=-\frac
{1}{\hbar}\operatorname*{Tr}X+\frac{1}{\hbar^{2}}X^{2}x\cdot x \label{tr}%
\end{equation}
hence the Fermi function of $\Psi_{X,Y}$ is the quadratic form
\begin{equation}
g_{\mathrm{F}}(x,p)=(p+Yx)^{2}+X^{2}x\cdot x-\hbar\operatorname*{Tr}X.
\label{gf3}%
\end{equation}
We can rewrite this formula as
\[
g_{\mathrm{F}}(x,p)=\left[  x\;,\;p\right]  M_{\mathrm{F}}%
\begin{bmatrix}
x\\
p
\end{bmatrix}
-\hbar\operatorname*{Tr}X
\]
where $M_{\mathrm{F}}$ is the symmetric matrix
\begin{equation}
M_{\mathrm{F}}=%
\begin{bmatrix}
X^{2}+Y^{2} & Y\\
Y & I
\end{bmatrix}
. \label{mf}%
\end{equation}
A straightforward calculation shows that%
\begin{equation}
M_{\mathrm{F}}=S^{T}%
\begin{bmatrix}
X & 0\\
0 & X
\end{bmatrix}
S \label{mfs}%
\end{equation}
where $S$ is the \emph{symplectic} matrix
\begin{equation}
S=%
\begin{bmatrix}
X^{1/2} & 0\\
X^{-1/2}Y & X^{-1/2}%
\end{bmatrix}
. \label{ess}%
\end{equation}

\section{Relation With the Wigner Function}

It turns out --and this is really a striking fact!-- that the matrix
(\ref{mfs}) is closely related to the Wigner transform
\begin{equation}
W\Psi_{X,Y}(z)=\left(  \frac{1}{2\pi\hbar}\right)  ^{n}\int_{\mathbb{R}^{n}%
}e^{-\frac{i}{\hbar}p\cdot y}\Psi_{X,Y}(x+\tfrac{1}{2}y)\Psi_{X,Y}^{\ast
}(x-\tfrac{1}{2}y)dy\label{oupsi}%
\end{equation}
of the state $\Psi_{X,Y}$ because we have%
\begin{equation}
W\Psi_{X,Y}(z)=\left(  \frac{1}{\pi\hbar}\right)  ^{n}\exp\left(  -\frac
{1}{\hbar}Gz\cdot z\right)  \label{goupsi}%
\end{equation}
where $G$ is the symplectic matrix%
\begin{equation}
G=S^{T}S=%
\begin{bmatrix}
X+YX^{-1}Y & YX^{-1}\\
X^{-1}Y & X^{-1}%
\end{bmatrix}
\label{G}%
\end{equation}
(see e.g. \cite{Birk,Littlejohn}). It follows from Eqn. (\ref{G}) that%
\begin{equation}
W\Psi_{X,Y}(z)=\left(  \frac{1}{\pi\hbar}\right)  ^{n}e^{-\operatorname*{Tr}%
X}\exp\left[  -\frac{1}{\hbar}g_{\mathrm{F}}(S^{-1}D^{-1/2}Sz)\right]
.\label{wgf}%
\end{equation}
with $D=%
\begin{bmatrix}
X & 0\\
0 & X
\end{bmatrix}
$. In particular, when $n=1$ and $\Psi_{X,Y}(x)=\Psi_{0}(x)$ the fiducial
coherent state (\ref{fid}) we have $S^{-1}D^{-1/2}S=I$ and $\operatorname*{Tr}%
X=1$ hence the formula%
\[
W\Psi_{0}(z)=\left(  \frac{1}{\pi\hbar}\right)  ^{1/4}\frac{1}{e}\exp\left[
-\frac{1}{\hbar}M_{\mathrm{F}}z\cdot z\right]
\]
which was already observed by Benenti and Strini \cite{best}.

\section{Geometric Interpretation}

In \cite{de02-2,de03-2,de04,de05,Birk,go09,degostat} (also see de Gosson and
Luef \cite{golu10}) we have applied the topological notion of symplectic
capacity \cite{Gromov,HZ,Polterovich} to the uncertainty principle. Recall
(ibid.) that the symplectic capacity $c(\mathcal{W})$ of an ellipsoid $Mz\cdot
z\leq1$ ($M$ a symmetric positive definite $2n\times2n$ matrix) is calculated
as follows: Consider the matrix product $JM$ ($J$ the standard symplectic
matrix); because $M$ is positive definite $JM$ is equivalent to the
antisymmetric matrix $M^{1/2}JM^{1/2}$ hence its $2n$ eigenvalues are of the
type $\pm i\lambda_{1}^{\sigma},..,$ $\pm i\lambda_{n}^{\sigma}$ where
$\lambda_{j}^{\sigma}>0$. The positive numbers $\lambda_{1}^{\sigma},..,$
$\lambda_{n}^{\sigma}$ are called the \emph{symplectic eigenvalues} of the
matrix $M$ and we have
\begin{equation}
c(\mathcal{W})=\pi/\lambda_{\max}^{\sigma}\label{cw}%
\end{equation}
where $\lambda_{\max}^{\sigma}=\max\{\lambda_{1}^{\sigma},..,$ $\lambda
_{n}^{\sigma}\}$. The symplectic capacity of a subset of phase space is
invariant under canonical transformations (linear or not).

We denote by $\mathcal{W}_{\mathrm{F}}$ the ellipsoid\ $M_{\mathrm{F}}z\cdot
z\leq\hbar\operatorname*{Tr}X$ bounded by the Fermi hypersurface
$\mathcal{H}_{\mathrm{F}}$ corresponding to the squeezed coherent state
$\Psi_{X,Y}$. Let us perform the symplectic change of variables $z^{\prime
}=Sz$; in the new coordinates the ellipsoid $\mathcal{W}_{\mathrm{F}}$ is
represented by the inequality
\begin{equation}
Xx^{\prime}\cdot x^{\prime}+Xp^{\prime}\cdot p^{\prime}\leq\hbar
\operatorname*{Tr}X\label{ferx}%
\end{equation}
hence $c(\mathcal{W}_{\mathrm{F}})$ is the symplectic capacity of the
ellipsoid (\ref{ferx}). Applying the rule above we thus have to find the
symplectic eigenvalues $D=%
\begin{bmatrix}
0 & X\\
-X & 0
\end{bmatrix}
$; a straightforward calculation shows that these are just the eigenvalues
$\lambda_{1},...,\lambda_{n}$ of $X$ and hence%
\begin{equation}
c(\mathcal{W}_{\mathrm{F}})=\frac{\pi\operatorname*{Tr}X}{\lambda_{\max}}%
\hbar\label{cwf}%
\end{equation}
where $\lambda_{\max}$ is the largest eigenvalue of $X$. Since $\lambda_{\max
}\leq\operatorname*{Tr}X\leq n\lambda_{\max}$ it follows that we have the
double inequality
\begin{equation}
\frac{1}{2}h\leq c(\mathcal{W}_{\mathrm{F}})\leq\frac{nh}{2}.\label{nh}%
\end{equation}
\  

A consequence of this is that the Fermi ellipsoid $\mathcal{W}_{\mathrm{F}}$
of a squeezed coherent state always contains a \textquotedblleft quantum
blob\textquotedblright, the image of a phase space ball $B^{2n}(z_{0}%
,\sqrt{\hbar}):|z-z_{0}|\leq\sqrt{\hbar}$ by a linear canonical transformation
(identified with a symplectic matrix $S$). A quantum blob is a phase space
ellipsoid with symplectic capacity $\pi\hbar=h/2$. The interest of quantum
blobs come from the fact that they provide us with a coarse-graining of phase
space different from the usual coarse graining by cubes with volume $\sim
h^{n}$ commonly used in statistical mechanics.

\section{Comments}

Benenti and Strini \cite{best} have given first order approximations
comparisons between of the level sets of the Wigner transform and the equation
$g_{F}(x,p)\leq0$ for sharply-peaked non-Gaussian quantum states; they claim
that the Fermi function can be used with profit for a semiclassical study of
such states. It would be very interesting to push their analysis further (and
in an arbitrary number of degrees of freedom). This can possibly be achieved
using known semiclassical approximations for the Wigner transform.

The quantity $\hbar^{2}\nabla_{x}^{2}R/R$ appearing in Eqns. (\ref{fermop})
and (\ref{gf2}) in the definitions of $\widehat{g}_{\mathrm{F}}$ and
$g_{\mathrm{F}}$ is a variant of the the quantum potential
\begin{equation}
Q=-\frac{\hbar^{2}}{2m}\frac{\nabla_{x}^{2}R}{R}\label{QP}%
\end{equation}
appearing in Bohmian mechanics (Bohm and Hiley \cite{BoHi}). It would be
interesting to interpret Fermi's function in terms of this popular variant of
quantum mechanics.

\end{document}